\documentclass[11pt,a4paper,twoside,groupcitations]{article}
\usepackage[T1]{fontenc}
\usepackage[ansinew]{inputenc}
\usepackage[english]{babel}
\usepackage{amsfonts}
\usepackage{amsmath}
\usepackage{bm}
\usepackage{array}
\usepackage{amsthm}
\usepackage{amssymb}
\usepackage{graphicx}
\usepackage{braket}
\usepackage{eucal}
\usepackage{verbatim}
\usepackage[table]{xcolor}
\usepackage{caption}
\usepackage{cite}
\usepackage{textcomp}
\usepackage{float}
\usepackage{multicol}
\usepackage{tikz}
\usetikzlibrary{positioning,arrows}
\usetikzlibrary{decorations.pathmorphing}
\usetikzlibrary{decorations.markings}
\usetikzlibrary{calc,decorations.markings}
\usetikzlibrary{arrows,shapes}
\usetikzlibrary{matrix,arrows}
\usepackage[colorlinks,hyperindex,unicode]{hyperref}
\definecolor{green1}{RGB}{0,128,0} 
\hypersetup{hidelinks,backref=true,pagebackref=true,hyperindex=true,colorlinks=true,breaklinks=true,urlcolor= blue}
\hypersetup{%
  colorlinks = true,
  linkcolor  = blue,
  citecolor = cyan,
}
\newcommand{\beq}{\begin{eqnarray}}
\newcommand{\eeq}{\end{eqnarray}}
\newcommand{\be}{\begin{eqnarray}}
\newcommand{\ee}{\end{eqnarray}}
\newcommand{\pro}[2]{\mbox{$\langle\, #1 \mid #2\,\rangle$}}

\renewcommand{\d}{\mbox{${\rm d}$}} 
\newcommand{\lp}{\ell_{\rm p}}
\newcommand{\mpl}{m_{\rm p}}
\newcommand{\gn}{G_{\rm N}}

\newcommand{\Rh}{R_{\rm H}}

\title{\bf Slowly rotating quantum dust cores and black holes}
\author{R.~Casadio$^{ab}$\thanks{E-mail: casadio@bo.infn.it}
$ $
and
L.~Tabarroni$^{a}$\thanks{E-mail: luca.tabarroni@studio.unibo.it}
\\
\\
$^a${\em Dipartimento di Fisica e Astronomia, Universit\`a di Bologna}
\\
{\em via Irnerio~46, 40126 Bologna, Italy}
\\
\\
$^b${\em I.N.F.N., Sezione di Bologna, I.S.~FLAG}
\\
{\em viale B.~Pichat~6/2, 40127 Bologna, Italy}
}
\begin{document}
\maketitle
\begin{abstract}
We study the effect of rotation on the spectrum of bound states for dust cores that source (quantum) black holes
found in Eur.~Phys.~J.~C 82 (2022) 10. 
The dust ball is assumed to spin rigidly with sufficiently slow angular velocity that perturbation
theory can be applied.
Like the total mass, the total angular momentum is shown to be quantised in Planck units, hence so is
the horizon area.
For sufficiently small fraction of mass in the outermost layer, the model admits ground states
which can spin fast enough within the perturbative regime so as to describe regular rotating objects
rather than black holes.
\end{abstract}
\section{Introduction}
\setcounter{equation}{0}
\label{S:intro}
Many proposals for resolving the classical spacetime singularities~\cite{HE} in approaches to quantum gravity have been
put forward (for a necessarily limited selection,
see Refs.~\cite{Hajicek:2001yd,Kuntz:2019lzq,Kuntz:2017pjd,Haggard:2014rza,Kuntz:2019gup,Piechocki:2020bfo}).
Quite remarkably, a general feature obtained in semiclassical
models of the gravitational collapse is the appearance of a bounce at a minimum radius~\cite{frolov,Casadio:1998yr,Schmitz:2020vdr},
which suggests that understanding the dynamics of matter in the description of black hole formation~\cite{Casadio:2016zpl}
(and subsequent evolution~\cite{Casadio:2019tfz}) is crucial. 
However, the non-linearity of Einstein's equations makes it impossible to study realistic classical models analytically,
and this furthermore renders their quantum description intractable in general.
\par
One can still hope to make some progress by studying those (over)simplified models obtained by forcing a
strong symmetry and unphysical equations of state for the collapsing matter which are employed to solve the Einstein
equations at the classical level.
One such example is given by the Oppenheimer and Snyder model of a ball of dust collapsing solely under
its own weight~\cite{OS}.
A discrete spectrum of bound states for this prototype of matter core was found in Ref.~\cite{Casadio:2021cbv}
(see also Ref.~\cite{Casadio:2022pla} for more results~\footnote{For thin shells,
see Refs.~\cite{vaz,hajicek,hk,Casadio:1995qy,Husain:2021ojz}.})
for which we will here estimate the effect of rotation.
The key idea in Ref.~\cite{Casadio:2021cbv} was to quantise the radial geodesic equation for dust located at the areal
radius of the ball in the Schwarzschild spacetime as an effective quantum mechanical description of the outermost layer. 
In order to deal with rotation, we shall here include the term in that geodesic equation containing the angular momentum
for a rigidly rotating ball and compute its perturbative effects on the spectrum for sufficiently slow angular velocity.
\section{Core quantum spectrum}
\setcounter{equation}{0}
\label{S:specttum}
Let us start by considering the collapse of a perfectly isotropic ball of dust with total ADM mass~\cite{adm} $M$
and areal radius $R=R(\tau)$.
Here $\tau$ is the proper time of the dust particles following radial geodesics $r=r(\tau)$ in the Schwarzschild
spacetime metric~\footnote{We shall always use units with $c=1$ and often write
the Planck constant $\hbar=\lp\,\mpl$ and the Newton constant $\gn=\lp/\mpl$, where $\lp$ and $\mpl$
are the Planck length and mass, respectively.}
\be
\d s^{2}
=
-\left(1-\frac{2\,\gn\,M_0}{r}\right)
\d t^{2}
+\left(1-\frac{2\,\gn\,M_0}{r}\right)^{-1}
\d r^{2}
+r^{2}
\left(\d\theta^{2}+\sin^{2}\theta\, \d\phi^{2}\right)
\ ,
\label{schw}
\ee
where $M_0=M_0(r)$ is the (constant) fraction of ADM mass inside the sphere of radius $r=r(\tau)$.
\par
In particular, we consider the outermost (thin) layer of (average) radius $r=R(\tau)$ and mass $\mu=\epsilon\,M$,
where $0<\epsilon<1$ is the fraction of mass in that layer.~\footnote{See Ref.~\cite{Casadio:2022pla} for more details
about the role of $\epsilon$.}
The evolution of $R$ is governed by the mass-shell condition for the massive layer
of four-velocity $u^\alpha=(\dot t,\dot R,0,0)$,
that is
\be
\label{geod-general}
\frac{E_{\mu}^{2}}{\mu^{2}}
-
\dot R^{2}
+
\frac{2\,\gn\,M_{0}}{R}
-
\left(1-\frac{2\,\gn\,M_{0}}{R}\right)
\frac{L_{\mu}^{2}}{R^{2}\,\mu^{2}}
=
1
\ ,
\ee
where dots denote derivatives with respect to $\tau$, $M_0=(1-\epsilon)\,M$, and $E_\mu$ and $L_\mu$ are
the conserved momenta conjugated to $t=t(\tau)$ and $\phi=\phi(\tau)$, respectively.
\subsection{Zero angular momentum and mass quantisation}
The case of purely radial motion was analysed in Refs.~\cite{Casadio:2021cbv,Casadio:2022pla},
which we briefly review here.
For $L_\mu=0$, Eq.~\eqref{geod-general} reads
\be
\label{geod-part}
H_0
\equiv
\frac{P^{2}}{2\,\epsilon\, M}
-\frac{\epsilon\left(1-\epsilon\right)\gn\,M^{2}}{R}
=
\frac{\epsilon\, M}{2}\left(\frac{E_\mu^{2}}{\epsilon^2\,M^{2}}-1\right)
\equiv
\mathcal E
\ ,
\ee
where $P=\epsilon\,M\,\dot R$ is the momentum conjugated to $R=R(\tau)$.
The canonical quantisation prescription with $\hat{P} =-i\,\hbar\,\partial_R$
allows us to write Eq.~\eqref{geod-part} as the time-independent Schr\"odinger equation
\be
\hat{H}_0\,\Psi_{\bar n}
=
\left[
-
\frac{\hbar^{2}}{2\,\epsilon\, M}
\left(
\frac{\d^2}{\d R^2}
+
\frac{2}{R}\,
\frac{\d}{\d R}
\right)
-
\frac{\epsilon\left(1-\epsilon\right)\gn\,M^{2}}{R}
\right]
\Psi_{\bar n}
=
\mathcal E_{\bar n}\,
\Psi_{\bar n}
\ ,
\ee
which is formally the same as the one for the $s$-states of the hydrogen atom.
The solutions are thus given by the Hamiltonian eigenfunctions
\be
\label{radial-wavefunction}
\Psi_{\bar{n}}(R)
=
\sqrt{\frac{\epsilon^{6}\left(1-\epsilon\right)^{3}M^{9}}{\pi\,\lp^{3}\,\mpl^{9}\,\bar{n}^{5}}}\,
\exp\!\left(-\frac{\epsilon^{2}\,(1-\epsilon)\,M^{3}\,R}{\bar{n}\,\mpl^{3}\,\lp}\right)
L_{\bar{n}-1}^{1}\!\!
\left(\frac{2\,\epsilon^{2}\,(1-\epsilon)\,M^{3}\,R}{\bar{n}\,\mpl^{3}\,\lp}\right)
\ ,
\ee
where the normalisation is defined in the scalar product
\be
\pro{\bar n}{\bar n'}
=
4\,\pi\,\int_0^\infty
R^2\,\Psi_{\bar n}^*(R)\,\Psi_{\bar n'}^{\phantom{}}(R)\,
\d R
=
\delta_{\bar n \bar n'}
\ ,
\ee
the functions $L_{{\bar n}-1}^1$ are Laguerre polynomials and the non-negative integer quantum number ${\bar n}$
corresponds to the eigenvalues 
\be
\mathcal E_{\bar n}
=
-
\frac{\epsilon^3\,(1-\epsilon)^2\,M^5}{2\,\bar n^2\,\mpl^4}
\ .
\label{EE_n}
\ee
The expectation value of the areal radius on these states is given by
\be
\bar R_{\bar n}
\equiv
\bra{\bar n} \hat R \ket{\bar n}
=
\frac{3\,\lp\,\bar n^2\,\mpl^3}{2\,\epsilon^2\,(1-\epsilon)\,M^3}
\ ,
\label{bRbn}
\ee
so that the quantum picture is the same that one would have in Newtonian physics.
In particular, the ground state $\bar n=1$ has a width $\bar R_1\sim \lp\,(\mpl/M)^3$ and energy $\mathcal E_1\sim -M\,(M/\mpl)^4$,
which makes it practically indistinguishable from a point-like singularity if $M\gg\mpl$.
\par
In fact, the only general relativistic feature that the model retains is given by $\mathcal E=\mathcal E(E_\mu)$
in Eq.~\eqref{geod-part}.
By then assuming that $E_\mu$ is well-defined for the allowed quantum states, we obtain
\be
\label{Emu}
0
\le
\frac{E_\mu^2}{\epsilon^2\,M^2}
=
1
-
\frac{\epsilon^2\,(1-\epsilon)^2}{\bar n^2}
\left(\frac{M}{\mpl}\right)^4
\ ,
\ee
which yields the lower bound 
\be
\bar n
\ge
N_M
\equiv
\epsilon\,(1-\epsilon)
\left(\frac{M}{\mpl}\right)^2
\ .
\label{N_M}
\ee
The fundamental state of the outermost layer hence corresponds to $\bar n=N_M\gg 1$, with
\be
\mathcal E_{N_M}
=
-\frac{\epsilon\,M}{2}
\ .
\ee
This result clearly resembles Bekenstein's famous area law quantisation~\cite{bekenstein},
since the allowed values of the mass and gravitational area are quantised in Planck units
according to~\footnote{We recall that $\Rh=2\,\gn\,M$ is the classical Schwarzschild (or gravitational) radius of the ball.}
\be
{\mathcal A}_M
=
16\,\pi\,\gn^2\,M^2
=
\frac{16\,\lp^2\,N_M}{\epsilon\,(1-\epsilon)}
\ ,
\label{A_M}
\ee
where the entire spectrum is given by $\bar n=N_M+n$, with $n$ a non-negative integer.
\par
We can now see that the singularity is precluded, as one expects from semiclassical
models~\cite{frolov,Casadio:1998yr}, since
\be
\bar R_{N_M}
=
\frac{3}{2}\,(1-\epsilon)\,\gn\,M
=
\frac{3}{4}\,(1-\epsilon)\,\Rh
\ ,
\ee
and $\bar R_{N_M}\sim \Rh\gg \lp$ unless the value of $\epsilon$ is extremely close to $1$.
Furthermore $\bar R_{N_M}<(3/4)\,\Rh$ for any value of $0<\epsilon<1$, so that a non-spinning
dust ball in the ground state can be the matter core of a quantum black hole for any mass $M$.
We also notice that the relative uncertainty in the areal radius is given by
\be
\frac{\Delta R_{\bar n}}{\bar R_{\bar n}}
\equiv
\frac{\sqrt{\bra{\bar n}\hat R^2\ket{\bar n}-\bar R_n^2}}{\bar R_{\bar n}}
=
\frac{\sqrt{\bar n^2+2}}{3\,\bar n}
\ ,
\label{DR/R}
\ee
which asymptotes to a minimum of $1/3$ for $\bar n\to \infty$.
In the following, for simplicity, we will mostly consider the ground state $\bar n=N_M$ with $M\gg\mpl$.~\footnote{For instance,
for $M=M_\odot\simeq 10^{30}\,$kg, one finds $N_M\sim 10^{76}$.}
\subsection{Rigidly rotating ball}
Next, we consider the case of a dust ball which rotates rigidly with angular velocity $\omega$.
A proper general relativistic treatment would require replacing the outer Schwarzschild geometry with the
Kerr metric, but we will here be satisfied with a perturbative approach for the quantum system described
previously.  
In particular, the outermost layer would have classical angular momentum
\be
L_\epsilon
=
\frac{2}{3}\,\epsilon\,M\,R^2\,\omega
\ ,
\ee
which we assume is small enough and replace for $L_\mu$ in Eq.~\eqref{geod-general}.
This yields the total Hamiltonian $H=H_0+V_L$, where $H_0$ is given in Eq.~\eqref{geod-part}
and
\be
V_L
&\!\!=\!\!&
\left(
1-\frac{2\,\gn\,(1-\epsilon)\,M)}{R}
\right)
\frac{L_\epsilon^2}{2\,\epsilon\,M\,R^2}
\nonumber
\\
&\!\!=\!\!&
\frac{2}{9}\,\epsilon\,M
\left[
R^2-2\,\gn\,(1-\epsilon)\,M\,R
\right]
\omega^2
\ .
\ee
In perturbation theory, this term will result in a correction to the eigenvalues~\eqref{EE_n} given by
\be
\Delta\mathcal E_{\bar n}
\equiv
\bra{\bar n}\,\hat V_L\,\ket{\bar n}
\simeq
\frac{2}{9}\,\epsilon\,M\,\bar R_{\bar n}
\left[
\frac{10}{9}\,\bar R_{\bar n}
-
2\,\gn\,(1-\epsilon)\,M
\right]
\omega^2
\ ,
\ee
where we used Eq.~\eqref{DR/R} for $\bar n\gg 1$ and $\bar R_{\bar n}$ is given in Eq.~\eqref{bRbn}.
\par
For the ground state $\bar n=N_M$, we thus find
\be
\Delta\mathcal E_{N_M}
&\!\!\simeq\!\!&
\frac{\mpl^4-\epsilon^2\,(1-\epsilon)^2\,M^4}{9\,\epsilon\,M}\,\gn^2\,\omega^2
\nonumber
\\
&\!\!\simeq\!\!&
-\frac{1}{9}\,\epsilon\,M\,(1-\epsilon)^2\,\gn^2\,M^2\,\omega^2
\ ,
\ee
in which we assumed that $\epsilon\,(1-\epsilon)\,M^2\gg\mpl^2$.
We then have
\be
\bra{N_M}\,\hat H\,\ket{N_M}
=
\mathcal E_{N_M}+\Delta \mathcal E_{N_M}
\simeq
\mathcal E_{N_M}
\left[
1
+
\frac{2}{9}\,(1-\epsilon)^2\,\gn^2\,M^2\,\omega^2 
\right]
\ .
\ee
This result is acceptable as long as $|\Delta \mathcal E_{N_M}|\ll |\mathcal E_{N_M}|$, or
\be
|\omega|
\ll
\frac{3}{\sqrt{2}\,(1-\epsilon)\,\gn\,M}
\equiv
\omega_{\rm max}
\ .
\label{omegac}
\ee
We further notice that $\omega_{\rm max}$ corresponds to a total angular momentum per unit mass
given by
\be
a_{\rm max}
\equiv
\frac{L_{\rm c}}{M}
=
\frac{2}{5}\,\bra{N_M}\,\hat R^2\,\ket{N_M}\,\omega_{\rm max}
\simeq
2\,(1-\epsilon)\,\gn\,M
\ .
\ee
\par
In the same approximation, we can compute the correction to the ground state quantum
number to leading order in $\omega$, which is given by
\be
N_{ML}
\simeq
N_M
\left[
1
+
\frac{1}{9}\,(1-\epsilon)^2\,\gn^2\,M^2\,\omega^2
\right]
\ .
\label{N_ML}
\ee
As expected, the angular momentum acts as an effective potential barrier and increases the
minimum allowed quantum number from $N_M$ to $N_{ML}$, corresponding to a larger radius
\be
\bar R_{N_{ML}}
\simeq
\bar R_{N_M}
\left[
1
+
\frac{2}{9}\,(1-\epsilon)^2\,\gn^2\,M^2\,\omega^2
\right]
\ .
\ee
In particular, for $\omega=\omega_{\rm max}$, one obtains
\be
N_{\rm max}
=
\frac{3}{2}\,\epsilon\,(1-\epsilon)\left(\frac{M}{\mpl}\right)^2
=
\frac{3}{2}\,N_M
\ ,
\ee
corresponding to a radius $\bar R_{\rm max}=2\,\bar R_{N_M}=3\,(1-\epsilon)\,\gn\,M$.
\subsection{Outer geometry}
\begin{figure}[t]
\centering
\includegraphics[width=7.8cm]{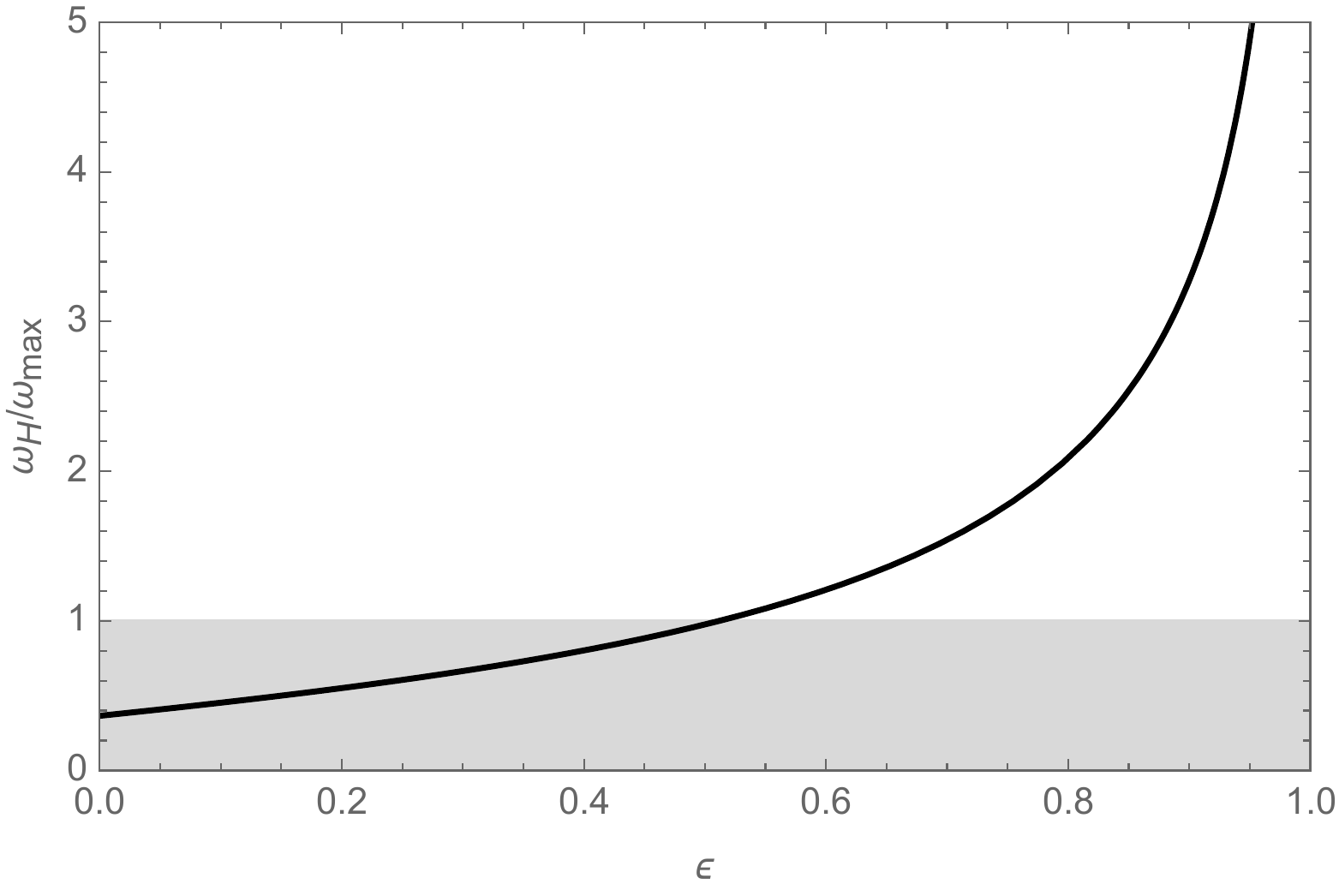}
$\ $
\includegraphics[width=8cm]{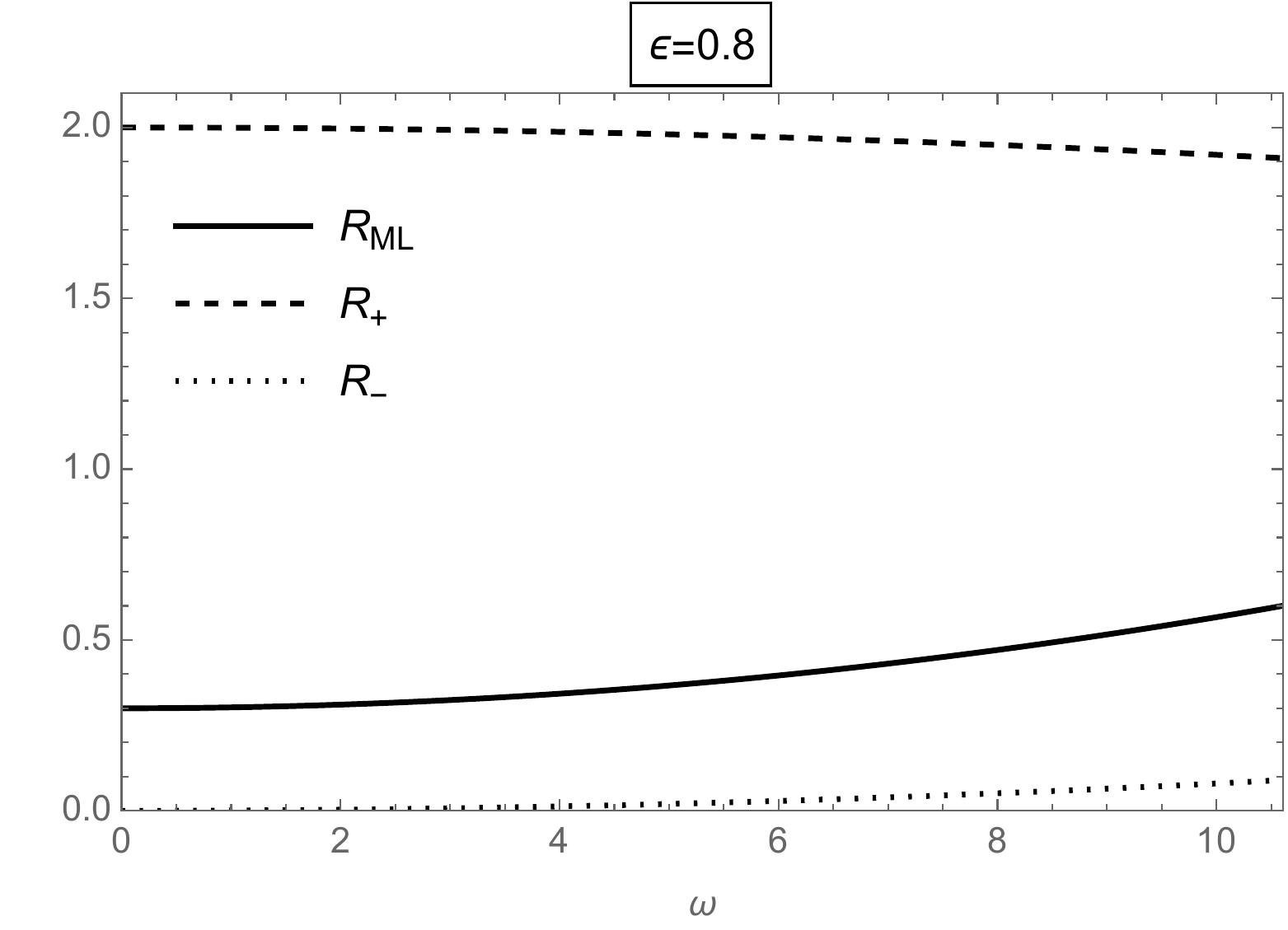}
\\
$ $
\\
\includegraphics[width=8cm]{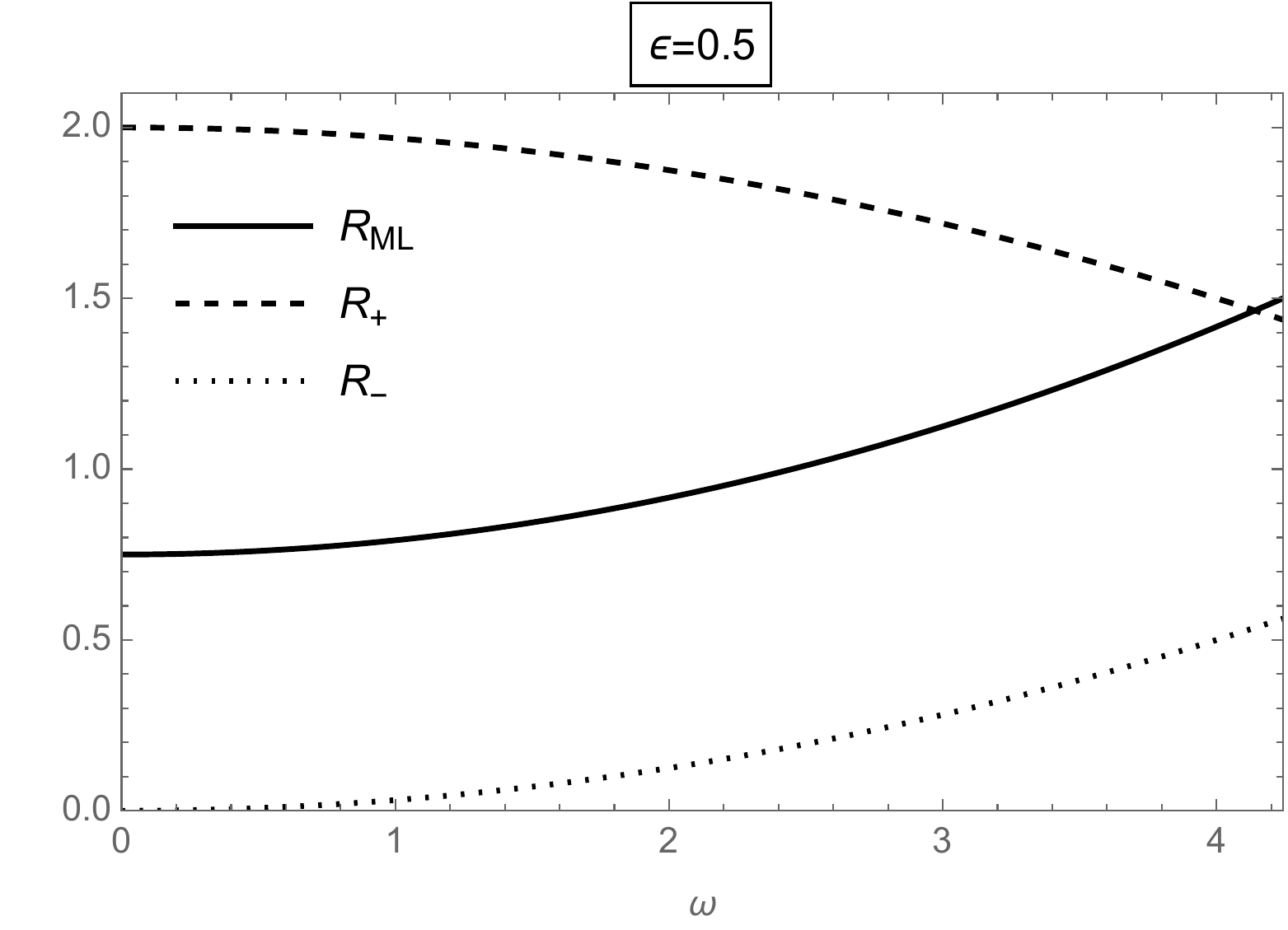}
$\ $
\includegraphics[width=8cm]{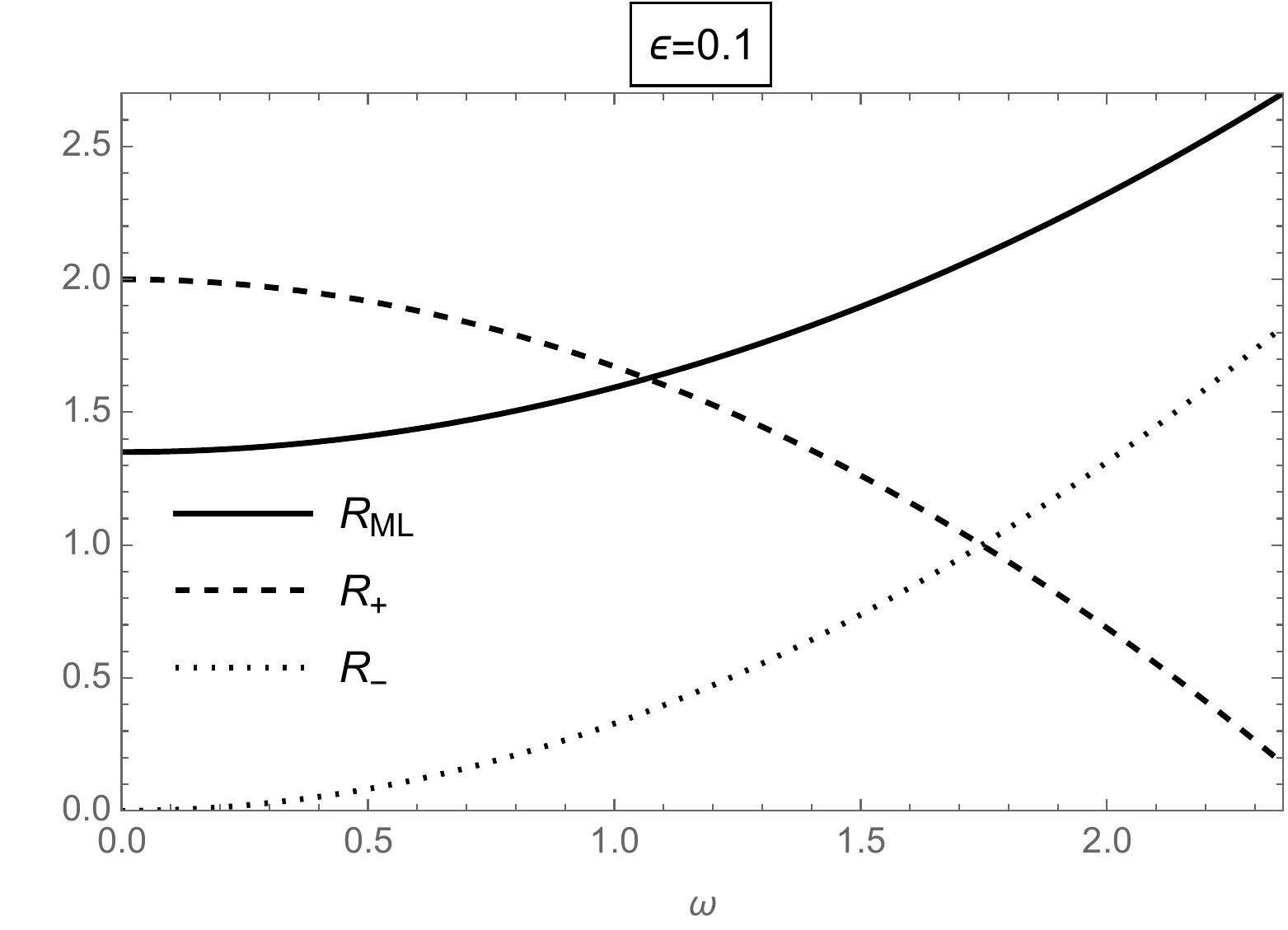}
\caption{Top left panel:
ratio $\omega_{\rm H}/\omega_{\rm max}$ for $0<\epsilon<1$ (shaded area corresponds to $\omega_{\rm H}<\omega_{\rm max}$).
Top right to bottom left panels: 
core radius (solid line), outer horizon radius (dashed line) and inner horizon radius (dotted line)
for $0<\omega<\omega_{\rm max}(\epsilon)$ with $\epsilon=0.8$, $0.5$ and $0.1$
(all quantities in units of $\gn\,M$).
}
\label{omegaR}
\end{figure}
We can now study the geometry outside the ball in the ground state by assuming that it is
indeed given by the Kerr metric.
We recall that the outer horizon for a Kerr metric is located at
\be
R_+
&\!\!=\!\!&
\gn\,M
+
\sqrt{\gn^2\,M^2-a^2}
\nonumber
\\
&\!\!\simeq\!\!&
2\,\gn\,M
\left[
1
-
\frac{1}{4}\,(1-\epsilon)^4\,\gn^2\,M^2\,\omega^2
\right]
\ ,
\label{R+}
\ee
whereas the inner Cauchy horizon would be at
\be
R_-
&\!\!=\!\!&
\gn\,M
-
\sqrt{\gn^2\,M^2-a^2}
\nonumber
\\
&\!\!\simeq\!\!&
\frac{1}{2}\,(1-\epsilon)^4\,\gn^3\,M^3\,\omega^2
\ll
\bar R_{N_M}
\ .
\ee
Since $R_-$ is located well inside the dust core, the relevant geometry will differ from the Kerr
vacuum there and the inner horizon cannot be realised (at least) within the perturbative approximation.
On the other hand, $R_+$ is larger than the core for 
\be
|\omega|
<
\omega_{\rm H}
=
\sqrt{\frac{3+9\,\epsilon}{(1-\epsilon)^3\,(5-3\,\epsilon)}}\,
\frac{1}{\gn\,M}
\ ,
\ee
which is approximately equal to $\omega_{\rm max}$ in Eq.~\eqref{omegac} for $\epsilon\simeq 0.5$
(see Fig.~\ref{omegaR}).
The inner matter core approaches the outer horizon radius $R_+$ for $\omega\simeq \omega_{\rm H}$,
with $\omega_{\rm H}<\omega_{\rm max}$ for $0<\epsilon\lesssim 0.5$ and
$\omega_{\rm H}>\omega_{\rm max}$ for $0.5\lesssim \epsilon<1$. 
In general, it would seem that one can consider small enough values of $\omega$ so that the outer horizon is realised.
Moreover, for sufficiently small $\epsilon$, one can also describe cores that spin fast enough,
so as not to form a black hole, within the perturbative regime.
We will come back to this point in the following.
\subsection{Angular momentum quantisation}
\begin{figure}[t]
\centering
\includegraphics[width=8cm]{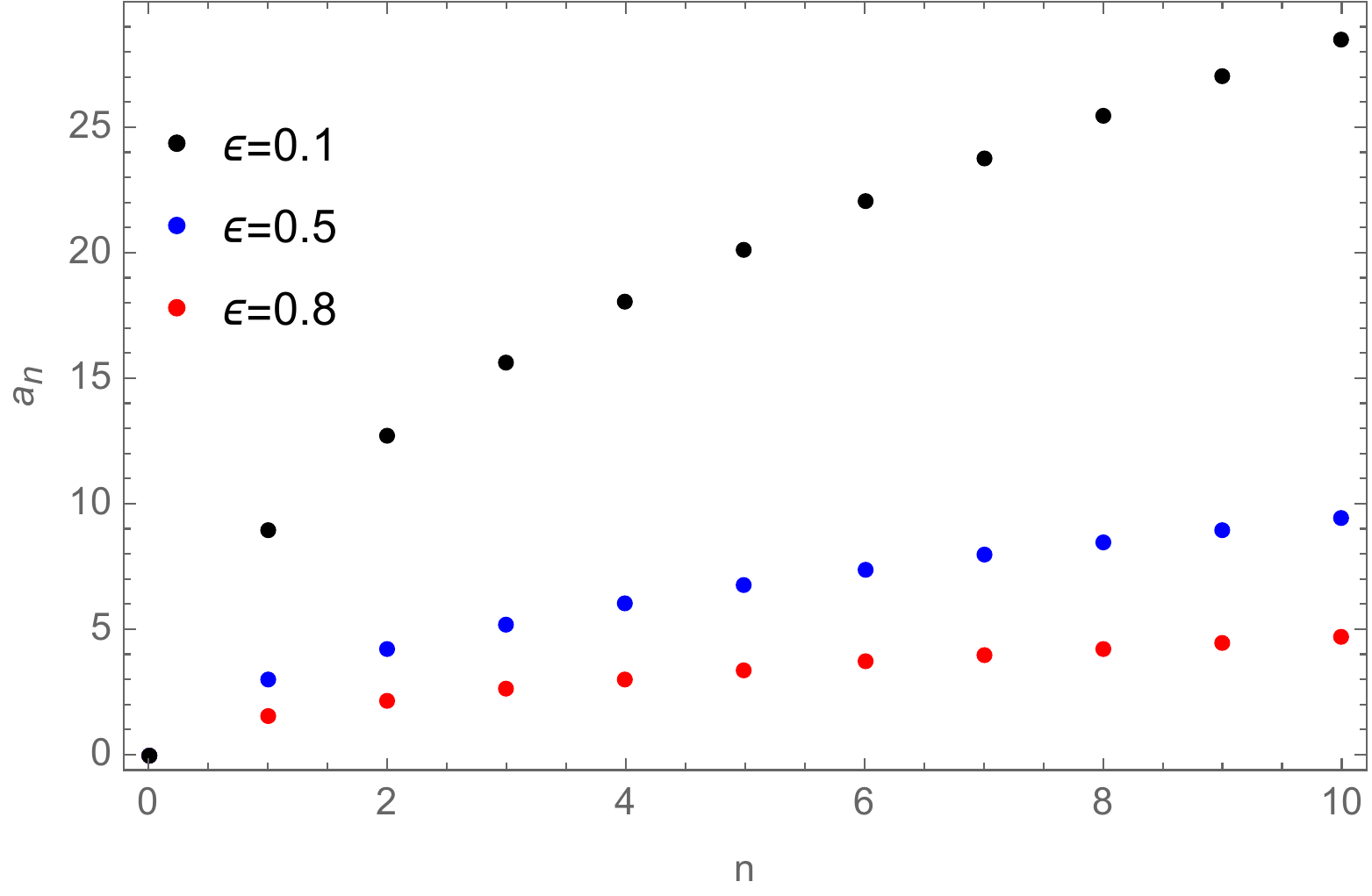}
$\ $
\includegraphics[width=8cm]{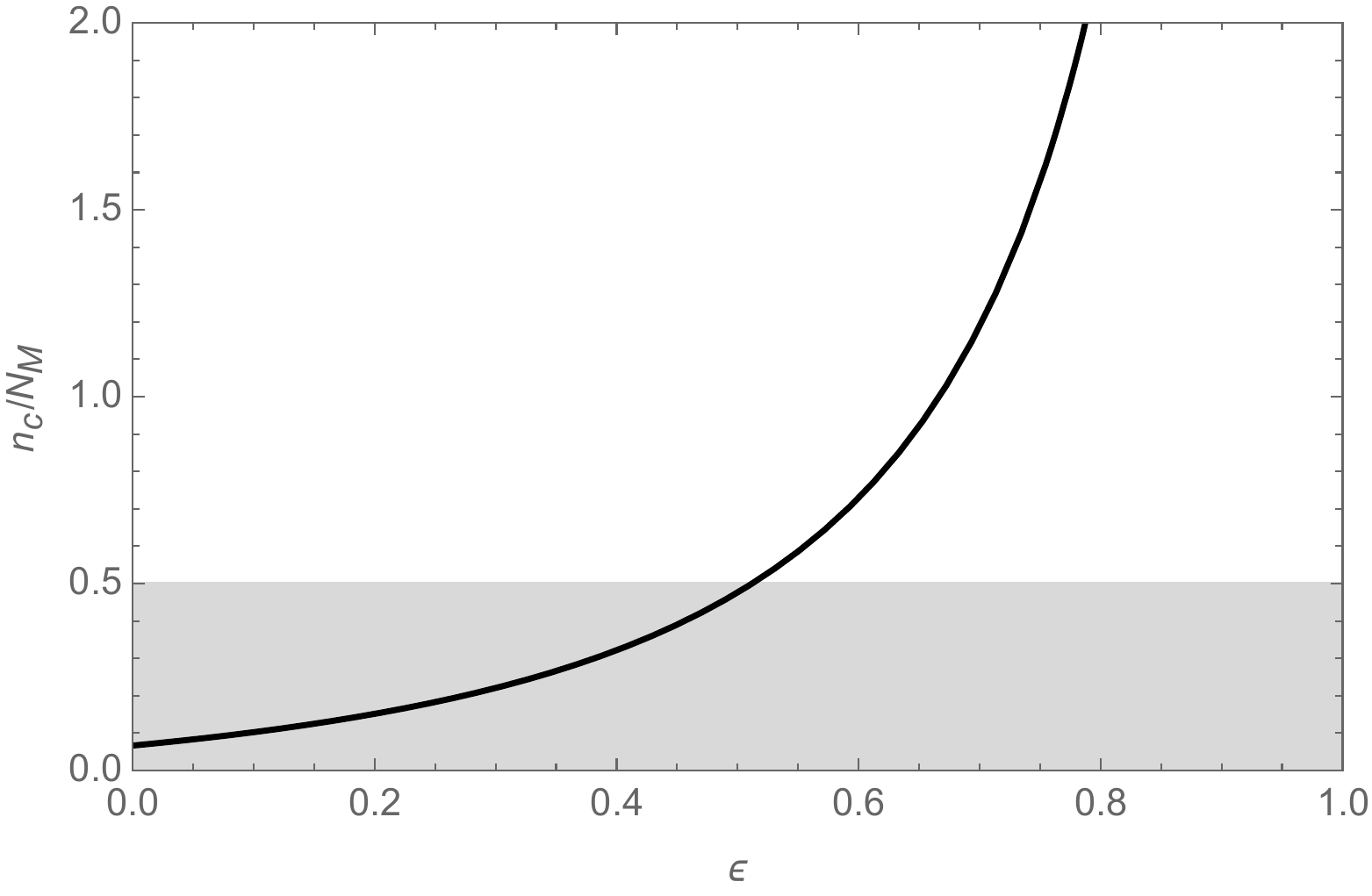}
\caption{Left panel: 
angular momentum-to-mass ratio~\eqref{a_n} (in units of $\lp$).
Right panel: ratio $n_{\rm c}/N_M$ in Eq.~\eqref{nc/NM} (shaded region marks the perturbative regime).
}
\label{an}
\end{figure}

In the above analysis, we considered the angular velocity $\omega$ as a (free) perturbative parameter,
but Eq.~\eqref{N_ML} implies the necessary quantisation of the angular momentum.
In fact, both $N_M$ and $N_{ML}$ must be (positive) integers in order to correspond to allowed
states in the spectrum~\eqref{radial-wavefunction}.
This implies that $N_{ML}=N_M+n$, with $n$ a (non-negative) integer (much smaller than $N_M$
for the perturbative result to hold), and we can write
\be
\omega^2
\simeq
\omega_n^2
=
\frac{9\,\epsilon\,n}{(1-\epsilon)\, N_M^2\,\lp^2}
\ ,
\ee
corresponding to a quantised angular momentum
\be
L_n
=
\frac{3}{\epsilon}\,\lp\,\mpl\,\sqrt{N_M\,n}
\ ,
\ee
or
\be
a_n
\equiv
\frac{L_n}{M}
=
3\,\lp\,\sqrt{\frac{1-\epsilon}{\epsilon}\,n}
\ ,
\label{a_n}
\ee
which is plotted in the left panel of Fig.~\ref{an} for a first few values of $n$ and the same values of $\epsilon$ used
in Fig.~\ref{omegaR}.
Note that the condition~\eqref{omegac} of small angular velocity then simply yields $n\ll N_M/2\sim N_M$.
\par
We can further notice that the quantisation of the mass and angular momentum lead to a quantisation
of the outer horizon radius~\eqref{R+} and area in Planck units, namely
\be
{\mathcal A}_+
=
4\,\pi\,R_+^2
\simeq
{\mathcal A}_M
\left[1-\frac{9}{2}\,(1-\epsilon)^2\,\frac{n}{N_M}
\right]
\ ,
\label{A+}
\ee
where ${\mathcal A}_M$ is the horizon area~\eqref{A_M} for a Schwarzschild black hole of mass $M$.
We then see that increasing the mass (that is $N_M$) makes the horizon larger, whereas increasing
the angular momentum (that is $n$) makes it smaller, consistently with the classical theory.
\section{Final remarks}
\setcounter{equation}{0}
\label{S:conc}
We have considered the perturbation induced by rotation on the quantum spectrum of dust balls found
in Ref.~\cite{Casadio:2021cbv}.
Quite expectedly, we have found that adding angular momentum increases the size of the ground state.
Moreover, no inner Cauchy horizon can appear within the perturbative regime, and the event horizon
area~\eqref{A+} is now quantised in terms of two quantum numbers, namely $\bar n=N_M$ corresponding
to the total mass $M$ and $n$ for the quantised total angular momentum of the system.
\par
Increasing $N_M$ makes both core and horizon larger, whereas increasing $n$ makes the core larger
but the event horizon smaller.
There is therefore a critical angular velocity $\omega_{n_{\rm c}}\simeq\omega_{\rm H}$, that is
\be
\frac{n_{\rm c}}{N_M}
\simeq
\frac{1+3\,\epsilon}{3\,(1-\epsilon)\,(5-3\,\epsilon)}
\ ,
\label{nc/NM}
\ee
above which the core remains larger than the (would-be outer) horizon and the system is not a black hole.
From the right panel of Fig.~\ref{an}, we see that this can occur within the perturbative regime if
$0<\epsilon\lesssim 0.5$, consistently with the top left panel of Fig.~\ref{omegaR}.
An example is provided by the bottom right panel of Fig.~\ref{omegaR}, where $\epsilon=0.1$
and the transition from a rotating black hole to a horizonless compact object occurs for $\gn\,M\,\omega\simeq 1$.
Of course, beside the qualitative behaviour, the precise values obtained from such a simple model should
not be trusted to bear a phenomenological relevance.
\section*{Acknowledgments}
R.C.~is partially supported by the INFN grant FLAG and his work has also been carried out in
the framework of activities of the National Group of Mathematical Physics (GNFM, INdAM).
%
%
%
%
%
%
\bibliographystyle{unsrt}

\end{document}